\documentclass[twocolumn,prb]{revtex4}
\usepackage[T1]{fontenc}
\usepackage{amsfonts}
\usepackage{amssymb}
\usepackage{times}
\usepackage{amsmath}
\usepackage{graphicx}

\let\mathbf=\boldsymbol
\def\beginABC{\begin{subequations}}
\def\endABC{\end{subequations}}

\begin{document}

\title{{\Large Symmetry Protected Topological Charge in Symmetry Broken
Phase:}\\
{\Large Spin-Chern, Spin-Valley-Chern and Mirror-Chern Numbers} }
\author{Motohiko Ezawa}
\affiliation{Department of Applied Physics, University of Tokyo, Hongo 7-3-1, 113-8656,
Japan }

\begin{abstract}
The Chern number is a genuine topological number. On the other hand, a
symmetry protected topological (SPT) charge is a topological number only
when a symmetry exists. We propose a formula for the SPT charge as a
derivative of the Chern number in terms of the Green function in such a way
that it is valid and related to the associated Hall current even when the
symmetry is broken. We estimate the amount of deviation from the quantized
value as a function of the strength of the broken symmetry. We present two
examples. First, we consider Dirac electrons with the spin-orbit coupling on
honeycomb lattice, where the SPT charges are given by the spin-Chern,
valley-Chern and spin-valley-Chern numbers. Though the spin-Chern charge is
not quantized in the presence of the Rashba coupling, the deviation is
estimated to be $10^{-7}$ in the case of silicene, a silicon cousin of
graphene. Second, we analyze the effect of the mirror-symmetry breaking of
the mirror-Chern number in a thin-film of topological crystalline insulator.
\end{abstract}

\maketitle

\section{Introduction}

There are two types of conserved charges, the N\"{o}ther charge and the
topological charge. The N\"{o}ther charge conserves as a result of the
equations of motion, while the topological charge conserves identically
irrespective of any local perturbation. The third type of conserved charges
may arise in the context of the symmetry protected topological (SPT) order.
The SPT order\cite{Gu,Turner,Chen, Oshikawa,Gu2} generates a topological
state only when some symmetry exists. The SPT charge is a topological charge
as far as the symmetry is not broken.

Topological insulator is one of the most fascinating concepts found in this
decade. It is characterized by topological numbers such as the Chern number%
\cite{TKNN} and SPT numbers. The Chern number is a genuine topological
charge, and indexes the quantum Hall state. Examples of SPT numbers are the $%
\mathbb{Z}_{2}$ index\cite{KaneMele} protected by the time-reversal symmetry
(TRS), the spin-Chern number\cite{Prodan1,LSheng,Sheng,Yang} protected by
the spin-rotation symmetry, and the mirror-Chern number\cite%
{TeoMirror,Takahashi} protected by the mirror symmetry. There are other SPT
numbers. Honeycomb systems are indexed by the valley-Chern and
spin-valley-Chern numbers in addition to the spin-Chern number due to the
valley-degree of freedom\cite{Fang11,Ezawa2Ferro}.

In this Letter we start with the Chern number $\mathcal{C}$ expressed in
terms of the single-particle Green function\cite{Volovik,Wang,Wang12,Gurarie}
rather than the single-particle Hamiltonian. A merit is that we can
calculate the Chern number even in the presence of interactions. We then
propose to define the SPT\ charge $\mathcal{C}_{\Lambda }$ corresponding to
a symmetry $\Lambda $ of the Hamiltonian in such a way that the Hall
conductivity $\sigma _{xy}^{\Lambda }$ is related to the SPT charge $%
\mathcal{C}_{\Lambda }$ by the well-known formula,%
\begin{equation}
\sigma _{xy}^{\Lambda }=\frac{e}{2\pi \hbar }\mathcal{C}_{\Lambda }.
\end{equation}%
The SPT charge $\mathcal{C}_{\Lambda }$ is expressed also in terms of the
single-particle Green function. It is a topological number and quantized
when the symmetry $\Lambda $ is unbroken. Our main result is that, even if $%
\Lambda $ is not a symmetry, the SPT charge $\mathcal{C}_{\Lambda }$ is well
defined and related to the Hall conductivity by this formula.

The SPT insulator may have symmetry protected gapless edge (surface) modes,
indicating the topological nature of this order\cite%
{Gu,Turner,Chen,Oshikawa,Gu2}. However, the SPT charge can be continuously
made zero through a continuous deformation of the Hamiltonian so as to break
the symmetry. The gapless edge (surface) modes disappear as a result of this
deformation though the gap of the bulk energy spectrum keeps open.

The SPT charge $\mathcal{C}_{\Lambda }$\ may be a continuous function of the
strength $\xi $ of the symmetry-breaking coupling. It is then shown that $%
\mathcal{C}_{\Lambda }=1-o(\xi ^{2})$. When the SPT charge is almost
quantized ($|\xi |\ll 1$), the associated Hall current is almost quantized.
The edge modes remains almost gapless.

We explicitly analyze Dirac electrons on honeycomb lattice in the presence
of the spin-orbit (SO) coupling. They form a QSH insulator, which is an SPT
state protected by TRS and by the spin $s_{z}$-symmetry. We consider a model
where TRS has been broken by the antiferromagnetic (AF) order $m_{z}$ in the
perpendicular direction. We then break the $s_{z}$-symmetry by introducing
the AF order in the in-plane component $m_{x}$. We calculate explicitly the
spin-Chern charge $\mathcal{C}_{s}$ as a function of these external
parameters. We show that $\mathcal{C}_{s}$ is continuously transformed from $%
\mathcal{C}_{s}=1$ to $\mathcal{C}_{s}=0$ by controlling them externally. We
also analyze the effect of the Rashba coupling on the spin-Chern charge, and
find $\mathcal{C}_{s}=1-5.9\times 10^{-7}$ in the case of silicene, the
silicon cousin of graphene. Consequently the spin-Chern charge is almost
quantized.

We also analyze a thin film of topological crystalline insulator\cite%
{Fu,Hsieh,LiuFu04,LiuFu}, which is protected by the mirror symmetry. The
mirror-Chern number is half quantized\cite{TeoMirror,Takahashi}. We
introduce a mirror-symmetry breaking term and calculate the mirror-Chern
charge. It becomes not quantized but the deviation is the second order in
the breaking-term strength.

\section{Chern number}

We analyze the two-dimensional system described by the Hamiltonian $H$ with
a gapped energy spectrum. The Chern number is an integral of the Berry
curvature over the first Brillouin zone\cite{TKNN}, 
\begin{equation}
\mathcal{C}=(2\pi )^{-1}\int d^{2}kF\left( \mathbf{k}\right) ,
\end{equation}%
where $F\left( \mathbf{k}\right) =\partial _{x}a_{y}-\partial _{y}a_{x}$ ($%
\partial _{x}\equiv \partial /\partial k_{x}$) and%
\begin{equation}
a_{k}=-i\sum_{\alpha }\left\langle \psi ^{\alpha }\left( \mathbf{k}\right)
\right\vert \partial _{k}\left\vert \psi ^{\alpha }\left( \mathbf{k}\right)
\right\rangle ,
\end{equation}%
with $\psi ^{\alpha }\left( \mathbf{k}\right) $ standing for the wave
function of the ground state which is an insulator. There exists an
alternative expression for the Chern number in terms of the Green function%
\cite{Ishikawa,Volovik}. With the use of the Matsubara Green function, 
\begin{equation}
G\left( \mathbf{k}\right) =[i\omega -H\left( \mathbf{k}\right) ]^{-1},
\end{equation}%
with $i\omega $ referring to the Matsubara frequency ($\omega $: real), the
Chern number is expressed as 
\begin{equation}
\mathcal{C}=\left( 2\pi \right) ^{-2}\int d^{2}k\int_{-\infty }^{\infty
}d\omega \,\Omega ,
\end{equation}%
with 
\begin{equation}
\Omega =\frac{1}{6}\varepsilon _{\mu \nu \rho }\text{Tr}[G\partial _{\mu
}G^{-1}G\partial _{\nu }G^{-1}G\partial _{\rho }G^{-1}],
\end{equation}%
where $k_{\mu }$, $k_{\nu }$ and $k_{\rho }$ run through $k_{0}\equiv
i\omega $, $k_{x}$ and $k_{y}$. This formula has a merit that it can be used
even in the presence of interactions\cite{Wang,Wang12,Gurarie}.

\section{Symmetry protected Chern number}

As far as the symmetry $\Lambda $ is not broken, the SPT charge $\mathcal{C}%
_{\Lambda }$ is a topological charge and quantized. An example is given by
the spin $s_{z}$-symmetry with $\Lambda =\sigma _{z}$, where the SPT charge
is the spin-Chern number $\mathcal{C}_{\Lambda }$ and observable by
measuring the Hall current,%
\begin{equation}
\sigma _{xy}^{\Lambda }=\frac{e}{2\pi \hbar }\mathcal{C}_{\Lambda }.
\label{HallCondu}
\end{equation}%
The main purpose of this work is to formulate the SPT charge $\mathcal{C}%
_{\Lambda }$ even when the symmetry $\Lambda $ is broken in such a way that
it is still related to the Hall current $\sigma _{xy}^{\Lambda }$ by formula
(\ref{HallCondu}).

We start with the SPT current $j_{x}$ defined by the symmetric product of
the velocity $v_{x}$ and the symmetry operator $\Lambda $ as a natural
extension of the spin current\cite{Sinova}, 
\begin{equation}
j_{x}=\frac{1}{2}\{v_{x},\Lambda \},  \label{current}
\end{equation}%
where $v_{i}=\hbar ^{-1}\partial _{i}H$. In the Kubo formalism the AC Hall
conductivity of the SPT charge is given by the correlation function between
the SPT current $j_{x}$ and the current $j_{y}=v_{y}$, 
\begin{align}
\sigma _{xy}^{\Lambda }\left( \varpi \right) =& \varepsilon _{xy\mu }\frac{e%
}{2\pi \hbar }K_{\mu }\left( \varpi \right) , \\
K_{\mu }\left( \varpi \right) =& \varepsilon _{\mu \alpha \beta }\frac{\hbar
^{2}}{\varpi }\int \frac{d^{2}k}{\left( 2\pi \right) ^{2}}\int_{-\infty
}^{\infty }d\omega  \notag \\
& \times \text{Tr}\left[ j_{\alpha }G\left( \omega +\varpi \right) j_{\beta
}G\left( \omega \right) \right] .
\end{align}%
The DC Hall conductivity is given by $\lim_{\varpi \rightarrow 0}\sigma
_{xy}^{\Lambda }\left( \varpi \right) $. By making the Taylor expansion of $%
G\left( \omega +\varpi \right) $ in $\varpi $, and using $\partial
_{i}H=\partial _{i}G^{-1}$, it is straightforward to derive%
\begin{equation}
\lim_{\varpi \rightarrow 0}K_{\mu }\left( \varpi \right) =\mathcal{C}%
_{\Lambda },
\end{equation}%
from which (\ref{HallCondu}) follows, where%
\begin{equation}
\mathcal{C}_{\Lambda }=\left( 2\pi \right) ^{-2}\int d^{2}k\int_{-\infty
}^{\infty }d\omega \,\Omega _{\Lambda },  \label{ChernGamma}
\end{equation}%
with%
\begin{equation}
\Omega _{\Lambda }=\frac{1}{6}\varepsilon _{\mu \upsilon \rho }\text{Tr}%
[G\Gamma _{\mu }G\Gamma _{\nu }G\Gamma _{\rho }],
\end{equation}%
and%
\begin{equation}
\Gamma _{x}=\frac{1}{2}\left\{ \Lambda ,\partial _{x}G^{-1}\right\} ,\quad
\Gamma _{y}=\partial _{y}G^{-1},\quad \Gamma _{0}=\partial _{0}G^{-1}.
\end{equation}%
In this derivation we have not assumed that $\Lambda $ is a symmetry, that
is, it can be that $\Lambda H\left( \mathbf{k}\right) \neq H\left( \mathbf{k}%
\right) \Lambda $. Consequently, for any symmetry $\Lambda $ for which the
formula (\ref{current}) makes sense, the SPT charge (\ref{ChernGamma}) is
valid and related to the Hall current (\ref{HallCondu}) even if the symmetry
is broken.

For the sake of completeness let us prove that, when $\Lambda $ is a
symmetry, that is, $\Lambda H\left( \mathbf{k}\right) =H\left( \mathbf{k}%
\right) \Lambda $, the SPT charge (\ref{ChernGamma}) is a conserved charge
independent of the equations of motion and quantized. A proof is made by
following Ref.\cite{Qi}. When the symmetry operator commutes with the Green
function $\Lambda G=G\Lambda $, $\Omega _{\Lambda }$ is simply given by%
\begin{equation}
\Omega _{\Lambda }=\frac{1}{6}\varepsilon _{\mu \nu \rho }\text{Tr}[\Lambda
G\partial _{\mu }G^{-1}G\partial _{\nu }G^{-1}G\partial _{\rho }G^{-1}].
\end{equation}%
We make a small modification $\delta H$ of the Hamiltonian $H$, which
results in the modification $\delta G$. By requiring that the perturbation
does not break the symmetry, or $\Lambda \delta G=\delta G\Lambda $, it is
straightforward to show that 
\begin{equation}
\delta \mathcal{C}_{\Lambda }=-\left( 2\pi \right) ^{-2}\int
d^{2}k\int_{-\infty }^{\infty }d\omega \,\delta \Omega ,
\end{equation}%
with 
\begin{equation}
\delta \Omega _{\Lambda }=\frac{1}{6}\varepsilon _{\mu \nu \rho }\partial
_{\mu }\text{Tr}[\Lambda G^{-1}\delta G\partial _{\nu }G^{-1}G\partial
_{\rho }G^{-1}G].
\end{equation}%
Since $\delta \mathcal{C}_{\Lambda }$ is an integral over a total
derivative, $\mathcal{C}_{\Lambda }$ is a topological charge. It is said to
be protected by the symmetry $\Lambda $. Such a conserved charge is the SPT
number.

When $\Lambda $ is an element of a Lie algebra, it is possible to identify $%
\Lambda $ with the diagonal element of the Pauli matrix $\sigma
_{z}^{\Lambda }$. Then the symmetry breaking coupling is given by $\xi
_{x}\sigma _{x}^{\Lambda }+\xi _{y}\sigma _{y}^{\Lambda }$ with external
parameters $\xi _{x}$ and $\xi _{y}$. By controlling them we may
continuously bring $\mathcal{C}_{\Lambda }$ to zero. Namely the $\Lambda $%
-SPT insulator with $\mathcal{C}_{\Lambda }=1$ is continuously transformed
into the $\Lambda $-trivial state with $\mathcal{C}_{\Lambda }=0$. Note that
the $\Lambda $-trivial state may still be an SPT state with respect to
another symmetry.

There exists an SPT charge $\mathcal{C}_{\Lambda }$ associated with each
symmetry $\Lambda $ of the unperturbated Hamiltonian density. We consider a
maximum set $\{\Lambda _{1},\Lambda _{2},\cdots ,\Lambda _{N}\}$ of these
symmetries each of which is commutative with another, $[\Lambda _{i},\Lambda
_{j}]=0$. It is interesting when the set may be identified with the Cartan
subalgebra of a certain Lie algebra. We can identify $\Lambda _{i}$ with the
diagonal element of the Pauli matrix $\sigma _{z}^{\Lambda _{i}}$. Then the
symmetry breaking coupling is given by $\xi _{x}^{i}\sigma _{x}^{\Lambda
_{i}}+\xi _{y}^{i}\sigma _{y}^{\Lambda _{i}}$ with external parameters $\xi
_{x}^{i}$ and $\xi _{y}^{i}$. The $\Lambda _{i}$-SPT state is continuously
transformed into the $\Lambda _{i}$-trivial state by controlling these
parameters. In the next section we shall see an example in the honeycomb
system where $\sigma _{z}$, $\eta _{z}$ and $\sigma _{z}\eta _{z}$
constitute the Cartan subalgebra of the SU(4) algebra with $\sigma _{z}$ and 
$\eta _{z}$ representing the spin and valley degrees of freedom.

In the rest of this paper we consider the system where we introduce a
symmetry-breaking perturbation term to the symmetric Hamiltonian. When it
contains a continuous parameter $\xi $, the Green function depends on $\xi $%
. We have $\partial \mathcal{C}_{\Lambda }/\partial \xi \equiv F(\xi )\neq 0$%
. By integrating it we find that $\mathcal{C}_{\Lambda }(\xi )$ changes
continuously from its quantized value as $\xi $ changes. We estimate $%
\mathcal{C}_{\Lambda }(\xi )$ for small symmetry-breaking perturbation.
Since $\Lambda $ is a symmetry without the perturbation we have $\mathcal{C}%
_{\Lambda }^{\prime }(0)=0$, or 
\begin{equation}
\mathcal{C}_{\Lambda }(\xi )=\mathcal{C}_{\Lambda }(0)+\frac{1}{2}\mathcal{C}%
_{\Lambda }^{\prime \prime }(0)\xi ^{2}+o(\xi ^{3}).
\end{equation}%
As far as perturbations are small ($|\xi |\ll 1$) the SPT charge $\mathcal{C}%
_{\Lambda }$ is almost quantized.

We discuss the bulk-edge correspondence associated with the SPT\ charge\cite%
{EzawaTaNaA}. When $\Lambda $ is a good symmetry ($\xi =0$), the SPT
insulator may have symmetry protected edge modes. However, as soon as the
symmetry is broken ($\xi \neq 0$), these gapless edge modes disappear though
the bulk gap keeps open. Nevertheless, as far as the symmetry breaking is
small ($|\xi |\ll 1$), we can use the emergence of almost gapless edge modes
as a signal of the SPT insulator. The edge-mode gap increases continuously
from zero as $|\xi |$ increases. The SPT charge $\mathcal{C}_{\Lambda }(\xi
) $ can be continuously transformed from the SPT phase ($\mathcal{C}%
_{\Lambda }=1$) to the trivial phase ($\mathcal{C}_{\Lambda }=0$). The
edge-mode gap becomes larger than the bulk gap in the trivial phase. See
Fig.1(a).

The emergence of gapless edge modes is a consequence of the reasoning that
the SPT\ number can change its quantized value discontinuously across the
edge only when the gap closes just as in the case of a genuine topological
number. Accordingly, no gapless edge modes would appear since there is no
need of gap closing provided the SPT number becomes ill-defined across the
edge, that is, in vacuum. We have already encountered such a case for the
valley-Chern and spin-valley-Chern numbers\cite{EzawaValley} because the
valley degree of freedom becomes ill-defined in vacuum. See also Section VI
and Fig.1.

\section{Dirac electrons on honeycomb lattice}

After describing general features of SPT charges we now present an example
to get an intuitive picture on them. Dirac electrons are ubiquitous in
monolayer honeycomb systems: There are four types of them corresponding to
the spin and valley degrees of freedom. It is intriguing that the SO
coupling makes Dirac electrons massive and turns the systems into
topological insulators\cite{KaneMele}. Silicene is a typical example\cite%
{LiuPRL}. It is particularly interesting since we can control various
topological phases externally by applying electric field\cite{EzawaNJP},
photo-irradiation\cite{EzawaPhoto} and exchange couplings\cite%
{EzawaQAHE,Ezawa2Ferro}.

The honeycomb lattice consists of two sublattices made of $A$ and $B$ sites.
The states near the Fermi energy are $\pi $ orbitals residing near the $K$
and $K^{\prime }$ points at opposite corners of the hexagonal Brillouin
zone. The low-energy dynamics in the $K$ and $K^{\prime }$ valleys is
described by the Dirac theory. In what follows we use notations $s_{z}=\pm 1$
for spin ($\uparrow ,\downarrow $), $\tau _{z}=\pm 1$ for sublattice ($A,B$)
and $\eta _{z}=\pm 1$ for valley ($K,K^{\prime }$). We also use the Pauli
matrices $\sigma _{a}$, $\tau _{a}$ and $\eta _{a}$ for the spin, the
sublattice pseudospin and the valley pseudospin, respectively.

We analyze the Dirac Hamiltonian\cite{KaneMele,Feng,Ezawa2Ferro,Hu}%
\begin{equation}
H_{0}^{\eta _{z}}=\hbar v_{\text{F}}\left( \eta _{z}k_{x}\tau _{x}+k_{y}\tau
_{y}\right) +\lambda _{\text{SO}}\eta _{z}\sigma _{z}\tau _{z}+m_{z}\sigma
_{z}\tau _{z}  \label{DiracHamil}
\end{equation}%
at the $K$ or $K^{\prime }$ point as the unpertubed symmetric Hamiltonian.
The first term represents electron hoppings with the Fermi velocity $v_{%
\text{F}}$. The second term describes the SO coupling\cite{KaneMele} with
coupling $\lambda _{\text{SO}}$. The third term has been introduced to break
TRS, whose physical meaning is the antiferromagnetic exchange magnetization%
\cite{Feng,Ezawa2Ferro,Hu} in the $z$-direction. There are three elements,%
\begin{equation}
\Lambda _{s}=\sigma _{z},\quad \Lambda _{v}=\eta _{z},\quad \Lambda
_{sv}=\sigma _{z}\eta _{z},  \label{SU4Cartan}
\end{equation}%
that commute with the Hamiltonian $H_{0}^{\eta _{z}}$, forming the Cartan
subalgebra of the SU(4) algebra. Accordingly we may introduce three SPT
numbers $\mathcal{C}_{s}$, $\mathcal{C}_{v}$ and $\mathcal{C}_{sv}$. They
are the spin-Chern\cite{Prodan1,Sheng,Sheng,Yang}, the valley-Chern\cite%
{Fang11,Fang13,Li} and the spin-valley-Chern numbers\cite{Fang11}.

They are calculable by using the Hamiltonian (\ref{DiracHamil}) for $H$ and
the operators (\ref{SU4Cartan}) for $\Lambda $ in (\ref{ChernGamma}), 
\begin{eqnarray}
\mathcal{C} &=&\mathcal{C}_{\uparrow }^{\text{K}}+\mathcal{C}_{\uparrow }^{%
\text{K'}}+\mathcal{C}_{\downarrow }^{\text{K}}+\mathcal{C}_{\downarrow }^{%
\text{K'}}, \\
\mathcal{C}_{s} &=&\frac{1}{2}(\mathcal{C}_{\uparrow }^{\text{K}}+\mathcal{C}%
_{\uparrow }^{\text{K'}}-\mathcal{C}_{\downarrow }^{\text{K}}-\mathcal{C}%
_{\downarrow }^{\text{K'}}), \\
\mathcal{C}_{v} &=&\mathcal{C}_{\uparrow }^{\text{K}}-\mathcal{C}_{\uparrow
}^{\text{K'}}+\mathcal{C}_{\downarrow }^{\text{K}}-\mathcal{C}_{\downarrow
}^{\text{K'}}, \\
\mathcal{C}_{sv} &=&\frac{1}{2}(\mathcal{C}_{\uparrow }^{\text{K}}-\mathcal{C%
}_{\uparrow }^{\text{K'}}-\mathcal{C}_{\downarrow }^{\text{K}}+\mathcal{C}%
_{\downarrow }^{\text{K'}}),
\end{eqnarray}%
where $\mathcal{C}_{s_{z}}^{\eta }=\frac{1}{2}{\eta _{z}}$sgn$(\Delta
_{s_{z}}^{\eta _{z}})$ with $\Delta _{s_{z}}^{\eta }$ being the Dirac mass, $%
\Delta _{s_{z}}^{\eta }=\eta _{z}s_{z}\lambda _{\text{SO}}+m_{z}s_{z}$.
There are two types of insulator phases in the model Hamiltonian (\ref%
{DiracHamil}). We find\cite{Ezawa2Ferro}%
\begin{equation}
(\mathcal{C},\mathcal{C}_{s},\mathcal{C}_{v},\mathcal{C}_{sv})=\left\{ 
\begin{array}{c}
(0,1,0,0)\text{ for }|m_{z}|<\lambda _{\text{SO}} \\ 
(0,0,0,1)\text{ for }|m_{z}|>\lambda _{\text{SO}}%
\end{array}%
\right. .  \label{CCspinNum}
\end{equation}%
We call the $(0,1,0,0)$ state a spin-Chern insulator for $m_{z}\neq 0$ since
it has a nontrivial spin-Chern number: It is protected by the $s_{z}$%
-symmetry. Recall that there is no TRS when $m_{z}\neq 0$. The system
undergoes a topological phase transition from the spin-Chern insulator ($%
\mathcal{C}_{s}=1$) to the trivial insulator ($\mathcal{C}_{s}=0$) as $m_{z}$
changes along the $m_{z}$ axis\cite{Ezawa2Ferro}. The phase transition
occurs at $m_{z}=\pm \lambda _{\text{SO}}$, where the gap closes. We remark
that this trivial insulator is actually the spin-valley-Chern insulator ($%
\mathcal{C}_{sv}=1$).

A comment is in order. TRS is recovered when we set $m_{z}=0$. In this case
the topological insulator is called the QSH insulator characterized by the $%
\mathbb{Z}_{2}$ index. When there exist both TRS and the $s_{z}$-symmetry,
the spin-Chern number is equal to the $\mathbb{Z}_{2}$ index mod$\,2$.

\section{Second Rashba term}

We first study how these numbers are affected when the Hamiltonian (\ref%
{DiracHamil}) is deformed to break the $s_{z}$-symmetry. The second Rashba
coupling exists in silicene as an intrinsic coupling\cite{LiuPRL,LiuPRB}. It
breaks the $s_{z}$-symmetry by mixing up and down spins on the \textit{%
next-nearest} neighbor hopping sites. The additional term\cite{LiuPRL,LiuPRB}
to the Dirac theory (\ref{DiracHamil}) is 
\begin{equation}
H_{\text{R2}}^{\eta _{z}}=a\lambda _{\text{R2}}\eta _{z}\tau _{z}\left(
k_{y}\sigma _{x}-k_{x}\sigma _{y}\right) ,  \label{2Rashba}
\end{equation}%
with $a$ the lattice constant. It is straightforward to calculate the SPT
charge (\ref{ChernGamma}) with $H^{\eta _{z}}=H_{0}^{\eta _{z}}+H_{\text{R2}%
}^{\eta _{z}}$ for the spin-Chern and spin-valley charges,\beginABC\label%
{MRashba}%
\begin{align}
\mathcal{C}_{s}& =\frac{\sum_{\eta _{z}}\eta _{z}\text{sgn}(\eta _{z}\lambda
_{\text{SO}}-m_{z})}{2\left[ 1+(a\lambda _{\text{R2}}/\hbar v_{\text{F}})^{2}%
\right] }, \\
\mathcal{C}_{sv}& =\frac{\sum_{\eta _{z}}\text{sgn}(\eta _{z}\lambda _{\text{%
SO}}-m_{z})}{2\left[ 1+(a\lambda _{\text{R2}}/\hbar v_{\text{F}})^{2}\right] 
}.
\end{align}%
\endABC This yields $\mathcal{C}_{s}=1-5.9\times 10^{-7}$, where we have
used $v_{\text{F}}=5.5\times 10^{5}$m/s, $a=3.86$\AA\ and $\lambda _{\text{R2%
}}=0.7$meV as sample parameters of silicene. Surely $\mathcal{C}_{s}$ does
not yield a quantized number, but the deviation of the spin-Chern charge
from 1 is negligibly small. Gapless edge modes must disappear when $%
m_{z}\neq 0$, but we do not recognize any discrepancy from zero within the
accuracy of numerical calculation. The spin mixing can be neglected in
practical purposes. We remark that gapless edge modes appear when $m_{z}=0$
however large $\lambda _{\text{R2}}$ may be, because they are protected by
TRS.

\section{In-plane AF order}

We next introduce the in-plane antiferromagnetic order $m_{x}$ to control
the $s_{z}$-symmetry breaking externally\cite%
{Rachel,WuRachel,Reuther,EzawaTaNaA} to the Dirac Hamiltonian (\ref%
{DiracHamil}). The additional term is 
\begin{equation}
H_{\text{AF}}^{\eta _{z}}=m_{x}\sigma _{x}\tau _{z}.  \label{HamilAF}
\end{equation}%
We neglect the second Rashba coupling (\ref{2Rashba}) since its effect is
negligible. The Hamiltonian$H^{\eta _{z}}=H_{0}^{\eta _{z}}+H_{\text{AF}%
}^{\eta _{z}}$ has the energy spectrum given by%
\begin{equation}
E\left( k\right) =\pm \sqrt{\left( \hbar v_{\text{F}}\right)
^{2}k^{2}+\left( \eta _{z}\lambda _{\text{SO}}-m_{z}\right) ^{2}+m_{x}^{2}},
\end{equation}%
with the band being $2|E\left( 0\right) |$. The system is an insulator
except for two isolated points $(\eta \lambda _{\text{SO}},0)$ with $\eta
_{z}=\pm 1$ in the $(m_{z},m_{x})$ phase diagram: See Fig.1(a). It is a
spin-Chern insulator ($\mathcal{C}_{s}=1,\mathcal{C}_{sv}=0$) in the region $%
(m_{z},0)$ with $|m_{z}|<\lambda _{\text{SO}}$, while it is a
spin-valley-Chern insulator ($\mathcal{C}_{s}=0,\mathcal{C}_{sv}=1$) for $%
|m_{z}|>\lambda _{\text{SO}}$.

\begin{figure}[t]
\centerline{\includegraphics[width=0.49\textwidth]{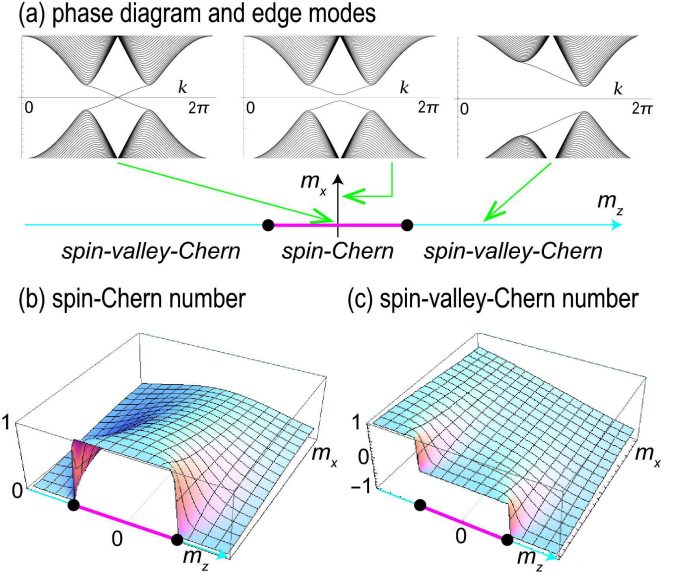}}
\caption{(a) Topological phase diagram of the Hamiltonain (\protect\ref%
{DiracHamil}) with (\protect\ref{HamilAF}) and edge-mode band structure of a
nanoribbon. The horizontal and vertical axes are the AF orders $m_{z}$ and $%
m_{x}$, respectively. The system is the spin-Chern insulator in the red
region with gapless edge modes, and in the spin-vallyer-Chern insulator in
the cyan region without gapless edge modes, on the $m_{z}$ axis. (b) The
spin-Chern charge $\mathcal{C}_{s}$ in the $(m_{z},m_{x})$ plane. (c) The
spin-valley-Chern charge $\mathcal{C}_{sv}$ in the $(m_{z},m_{x})$ plane.
They are quantized and represents topological quantum numbers only along the 
$m_{z}$ axis. They are continuous functions in all other region.}
\label{FigInAF}
\end{figure}

We place a zigzag nanoribbon in vacuum: See Fig.1(a). We find that gapless
edge modes emerge only in this spin-Chern insulator region\cite{EzawaTaNaA}.
This is because the spin-Chern number is well-defined but the
spin-valley-Chern number is not for the vacuum\cite{EzawaValley}. The
spin-Chern number cannot change its value from $\mathcal{C}_{s}=1$ to $%
\mathcal{C}_{s}=0$ across the edge without gap closing. On the other hand,
there is no need of the gap closing since the spin-valley-Chern number is
not defined in the vacuum.

We use (\ref{ChernGamma}) with $H^{\eta _{z}}=H_{0}^{\eta _{z}}+H_{\text{AF}%
}^{\eta _{z}}$ to calculate the spin-Chern and spin-valley-Chern charges,%
\beginABC\label{ChernAF}%
\begin{align}
\mathcal{C}_{s}& =\sum_{\eta _{z}=\pm 1}\frac{\lambda _{\text{SO}}-\eta
_{z}m_{z}}{2\sqrt{\left( \eta _{z}\lambda _{\text{SO}}-m_{z}\right)
^{2}+m_{x}^{2}}}, \\
\mathcal{C}_{sv}& =\sum_{\eta _{z}=\pm 1}\frac{\eta _{z}\lambda _{\text{SO}%
}-m_{z}}{2\sqrt{\left( \eta _{z}\lambda _{\text{SO}}-m_{z}\right)
^{2}+m_{x}^{2}}}.
\end{align}%
\endABC They are quantized only when $m_{x}=0$, yielding (\ref{CCspinNum}).
They are continuous function in all other region: See Fig.1(b) and (c).
However, we may control parameters $m_{x}$ and $m_{z}$ so that the phase
transition takes place without gap closing. Indeed, we can choose a path
connecting the spin-Chern insulator ($\mathcal{C}_{s}=1,\mathcal{C}_{sv}=0$)
and the spin-valley-Chern insulator ($\mathcal{C}_{s}=0,\mathcal{C}_{sv}=1$)
along which the gap never closes, $|E\left( 0\right) |\neq 0$, and the
spin-Chern and spin-valley-Chern charges continuously change.

\section{First Rashba term}

We also analyze the effect from the first Rashba coupling, which is an
essential ingredient of the Kane-Mele model\cite{KaneMele}. The coupling
mixes up and down spins on the \textit{nearest} neighbor hopping sites. It
yields the following term to the Dirac Hamiltonian (\ref{DiracHamil}),%
\begin{equation}
H_{\text{R1}}=\frac{1}{2}\lambda _{\text{R1}}\sum_{\eta _{z}}(\eta _{z}\tau
_{x}\sigma _{y}-\tau _{y}\sigma _{x}).
\end{equation}%
We calculate the SPT charge (\ref{ChernGamma}) for the spin-Chern and
spin-valley-Chern charges with the use of the SO term and the first Rashba
term, \beginABC%
\begin{align}
\mathcal{C}_{s}& =\frac{\sum_{\eta _{z}}\eta _{z}\text{sgn}(\eta _{z}\lambda
_{\text{SO}}-m_{z})}{2}\frac{4-\xi ^{-1}\text{arctanh}\xi }{3}, \\
\mathcal{C}_{sv}& =\frac{\sum_{\eta _{z}}\text{sgn}(\eta _{z}\lambda _{\text{%
SO}}-m_{z})}{2}\frac{4-\xi ^{-1}\text{arctanh}\xi }{3},
\end{align}%
\endABC where $\xi =|\lambda _{\text{R1}}/\lambda _{\text{SO}}|$.

\section{Topological crystalline insulator thin film}

We next study the topological crystalline insulator (TCI), which is a new
type of topological insulator protected by the crystal group symmetry. It
has been predicted\cite{Hsieh} that the TCI is realized in three-dimensional
materials, SnTe and Pb$_{1-x}$Sn$_{x}$Se(Te). Gapless surface states have
been observed\cite{Tanaka,Dziawa,Xu} in these materials. In particular,
their surface system is characterized by the mirror-Chern number\cite%
{Fu,Hsieh,LiuFu04,LiuFu}.

Dirac electrons emerge on the surface. We consider a thin film made of a
TCI, where there are interferences between the two surfaces. The effective
Hamiltonian for a TCI thin film is given by a two-dimensional model\cite%
{LiuFu}, 
\begin{equation}
H(k)=(v_{x}k_{x}\sigma _{x}-v_{y}k_{y}\sigma _{y})\tau _{x}+m\tau _{z},
\end{equation}%
where $\sigma _{i}$ is the Pauli matrix of spins and $\tau _{i}$ is that of
pseudospins representing two Dirac cones existing in the bulk TCI. Note that
there are two Dirac cones in the bulk TCI, which are projected into one
Dirac cone with the same momentum on the [001] TCI surface. The velocities $%
v_{x},v_{y}$ and the Dirac mass $m$ are to be derived from microscopic
parameters of surface states in the 3D TCI and their hybridization strengths.

This Hamiltonian has a mirror symmetry $\Lambda _{M}H(k)\Lambda
_{M}^{-1}=H(k)$ with 
\begin{equation}
\Lambda _{M}=-i\sigma _{z}\tau _{z}.
\end{equation}
The energy spectrum is given by%
\begin{equation}
E=\pm \sqrt{v_{x}^{2}k_{x}^{2}+v_{y}^{2}k_{y}^{2}+m^{2}}.
\end{equation}%
The mirror-Chern number is calculated from (\ref{ChernGamma}) as 
\begin{equation}
\mathcal{C}_{M}=\frac{1}{2}\text{sgn}(mv_{x}v_{y}).
\end{equation}%
It has two phases characterized by $\mathcal{C}_{M}=\pm \frac{1}{2}$. The
band is inverted and the system is topological for $m<0$, while it is
trivial for $m>0$. A topological phase transition occurs at $m=0$ with gap
closing.

We break the mirror symmetry by introducing the term $H^{\prime }=\lambda
_{E}\tau _{y}$, satisfying $\Lambda _{M}H^{\prime }\Lambda
_{M}^{-1}=-H^{\prime }$. The energy spectrum is modified as%
\begin{equation}
E=\pm \sqrt{v_{x}^{2}k_{x}^{2}+v_{y}^{2}k_{y}^{2}+m^{2}+\lambda _{E}^{2}}.
\end{equation}%
The gap closes only at one point ($m=0$, $\lambda _{E}=0$). The mirror-Chern
charge is calculated as 
\begin{equation}
\mathcal{C}_{M}=\frac{1}{2}\frac{m}{\sqrt{m^{2}+\lambda _{E}^{2}}}\text{sgn}%
(v_{x}v_{y}).
\end{equation}%
It changes continuously from the topological insulator with $\mathcal{C}%
_{M}=-\frac{1}{2}$ ($m<0$, $\lambda _{E}=0$) to the topological insulator
with $\mathcal{C}_{M}=\frac{1}{2}$ ($m>0$, $\lambda _{E}=0$) without gap
closing by first switching on $\lambda _{E}$, then changing $m$ and finally
switching off $\lambda _{E}$. The mirror-Chern charge is no longer quantized
in the presence of the mirror-symmetry breaking term, but its deviation is
the second order in $\lambda _{E}$.

\section{Discussions}

The SPT charge is a topological number and quantized when the associated
symmetry is unbroken. We have presented its Green function representation
valid even in the symmetry-broken phase. Our main result is that, even if $%
\Lambda $ is not a symmetry, the SPT charge (\ref{ChernGamma}) is well
defined and related to the Hall current formula (\ref{HallCondu}). For
instance, the spin current is observed as the difference between the up-spin
and down-spin currents with respect to the $z$ axis even if the spin is not
a good symmetry. It is interesting that our formulas are valid even if the
symmetry breaking is large. For instance, the spin-Chern charge (\ref%
{MRashba}) or (\ref{ChernAF}) is valid even for any value of $\lambda _{%
\text{R2}}$ or $m_{x}$. The spin-Chern charge simply becomes zero when they
are large enough. It implies that the average value of the $s_{z}$ component
becomes zero, which is physically reasonable.


We have elsewhere proposed possible topological devices\cite%
{EzawaAPL13,EzawaValley} with the use of edge channels carrying SPT charges
in a honeycomb system. Provided the edge-mode gap is far less than
experimental resolution, we may use these almost gapless edge modes as
carriers of currents. In principle, pure samples can be fabricated, where
the effects of symmetry-breaking impurities must be made negligible. Then,
SPT charges may well be treated as if they were topological numbers.

I am very much grateful to N. Nagaosa, Y. Tanaka, S. Rachel and R. Takahashi
for many fruitful discussions on the subject. This work was supported in
part by Grants-in-Aid for Scientific Research from the Ministry of
Education, Science, Sports and Culture No. 25400317.

\end{document}